\documentclass{article}
\usepackage{PRIMEarxiv}
\usepackage[utf8]{inputenc}  
\usepackage[T1]{fontenc}     
\usepackage{hyperref}        
\usepackage{url}            
\usepackage{booktabs}       
\usepackage{amsfonts}    
\usepackage{nicefrac}     
\usepackage{microtype}     
\usepackage{lipsum}		 
\usepackage{graphicx}
 \usepackage{float}
\usepackage{doi}
\usepackage{amsthm}
\usepackage{amsmath}
\usepackage{amssymb}
 
\usepackage{orcidlink}
\title{Speedrunning and path integrals}
\author{Gabriele Lami  \,\orcidlink{0000-0002-2303-2967}  \\ gabriele.lami@tutanota.com}  
\date{ 2024}

\begin{document}
\maketitle
\tableofcontents
\section{Introduction}
 
The purpose of this article is to draw an analogy between the speedrunning community and the principles of quantum mechanics. While classical physics is the foundation of our understanding of the physical world, it is not the only way to approach reality. In this article, we will explore the concept of speedrunning as a representation of a simplified version of quantum mechanics within a classical simulation. This analogy can be seen as a simplified approach to understanding the broader idea that quantum mechanics may emerge from classical mechanics simulations (in an implementation of a Turing machine\footnote{For a classic introduction to the subject see, for example, \cite{marvin}. For the purposes of this article, a universal Turing machine is not required because it is not necessary to have a simulator that can exactly simulate itself.}) due to the limitations of the simulation.\\
The concept of speedrunning will be explored from the perspective inside the simulation, where the player is seen as a ``force of nature'' that can be interpreted through Newton's first law.\footnote{
``Every body persists in its state of being at rest or of moving uniformly straight forward, except insofar as it is compelled to change its state by force impressed.''
%``Corpus omne perseverare in statu suo quiescendi vel movendi uniformiter in directum, nisi quatenus à viribus impressis cogitur statum illum mutare.''
Philosophiae Naturalis Principia Mathematica, Is. Newton, Londini, iussu Societatis Regiae ac typis Josephi Streater, anno MDCLXXXVII.}\\
Starting from this general assumption, the aim is to build a bridge between these two fields by using the mathematical representation of path integrals. The midpoint will be the explanation of how, if represented correctly, the activity of an expert player of a video game, in the act of performing a series of speedrunning attempts, can be seen as an agent in a pseudo-physics simulation tending to minimize a certain functional. This can be seen as the implementation of a stationary action principle \cite{stationary}.
The use of the stationary action principle as a surrogate for the variational representation of classical physics provides a direct link to our goal, the path integral introduced by Feynman.\\ 
In theoretical physics, the path integral formulation makes it possible to calculate the probability of the outcome of an interaction between quantum objects and fields, using a specific type of coefficient relative to all the different possible paths. In the act of speedrunning, the player actively explores various possible paths within the game or simulation, allowing for the implementation of ideal but imperfect decisions in the pseudo-physical world. This can be interpreted as a sum over the paths, considering the probability of generating an ideal path in the context of the sum of all possible paths.
Since a real player plays a certain number of games, it will also be explored how probability in speedrunning can be interpreted in a pseudo-``many worlds theory'' perspective. 
In this context, the alternative worlds do not coexist in the same universe and are then separated in the case of measurement, but are sequential in a time outside the universe under consideration (the simulation), which in this case is relative to the player's physical space.
This kind of approach also provides a simplified but effective tool for reasoning about the concept represented by the simulation hypothesis.\footnote{See, for example, \cite{nick}.}   In the case of a real video game, the computer code of a physical simulation is actually available and can be interpreted within the space in which it is played.\\
A use of such an approach as an intermediate layer between machine learning techniques aimed at finding an optimal strategy and a game simulation is also analysed.\\
This article will focus primarily on the relationship between classical and quantum physics within the simulation, leaving aside more technical (in this context) issues in field theory such as invariance with respect to Lorentz transformations and virtual particles. While these concepts are certainly important in the field of physics, they will not be explored in depth in this article.
It is our intention to develop these themes in later writings, as they are of considerable value in extending the proposed analogy.

\section{Speedrunning and the  stationary principle}
Speedrunning is a type of competitive game whose goal is for the players to complete a video game in the shortest possible time.
The value to be optimized is, therefore, the running time of the game, but there are different categories of competition with different constraints that make the objective more complex.
Our intent here is to examine the subset of video games known as platformer games, because it is easier to present our approach in this particular context.
A prototypical example, given its long history in the context of competitive gaming, is Super Mario Bros.\footnote{The game was developed and published by Nintendo for the Nintendo Entertainment System (NES) in 1985.} (SMB1), which I will use as a reference.
This choice is based on both the iconic status of the game in the speedrunning community and the interest it has aroused over the years in various academic fields, from computational complexity and AI to aesthetics.\footnote{A few examples are \cite{tricky} \cite{baroque} \cite{superai} \cite{lstm} \cite{accidents} \cite{tas}.}\\ 
In SMB1, for example, the goal might be to complete the game in the shortest possible time while collecting a minimum number of items or completing a minimum number of stages. Two important categories common to many games are any\% and 100\%. Here the percentage indicates the completion value for the specific game; for example, in SMB1, 100\% means completing all levels.
In a game like SMB1 whose pseudo-random components are limited, the professional player is able to achieve a stable percentage of runs around a given average value of game time. The record (as of 2023-09-06) for the category any\% is 4m 54s 631ms and is held by Niftski.\footnote{https://www.speedrun.com/smb1} This record is particularly meaningful because it equals the time that is considered the minimum to complete all but the last screen with the known best strategies. In this case, the theoretical value of the minimum is assumed using empirical evaluation methods.\\
In our context, it is also important to briefly analyze how a player habitually achieves a record.
In general, speedrunning rankings are based on game plays made with certain rules to prevent fraud. If a player appears to achieve a record, their performance is analyzed and validated.
Two conditions required by most ranking entities are the recording of the screen and, for some games, of the player's hands with two different cameras.\footnote{https://www.speedrunslive.com/rules-faq/faq} In general, the rankings are independent of specific events such as fairs or championships, although there are major events, such as Games Done Quick (GDQ) \cite{gdq}.The records are validated by entities accepted by the community, like https://www.speedrun.com/smb1 .
Many speedrunners try to achieve a record with consecutive runs in sessions played while streaming live on online platforms.\\
This kind of ``performance'' is interesting to analyze from the perspective of a physical system.
In many games, including SMB1, the virtual representative of the player within the simulation, the avatar,\footnote{There is substantial literature, including in academic publications, on the concept of avatars in gaming and in the virtual context in general. See, for example, \cite{avatar}.} follows a pattern in the pseudo-physical simulation represented by the game. In SMB1, the game takes place in a two-dimensional reality with simplified concepts of gravity and collisions.
Each run, then, can be seen as the trajectory of the avatar in simulation time.
It is also important to note that each run can be compared with other runs. Individual runs are replays of the same scenario starting at the same instant $t$ within simulation time. The start time of each run can always be taken as the origin of the time axis, and internal time is quantized due to the finiteness of the simulator's memory.\\
When analyzing a streaming session of a professional player, it is possible to see a set of neighboring avatar trajectories in the space of simulation states, covering most of the total number of runs.
 \begin{figure}[H]
    \centering
    \includegraphics[width=0.5\linewidth]{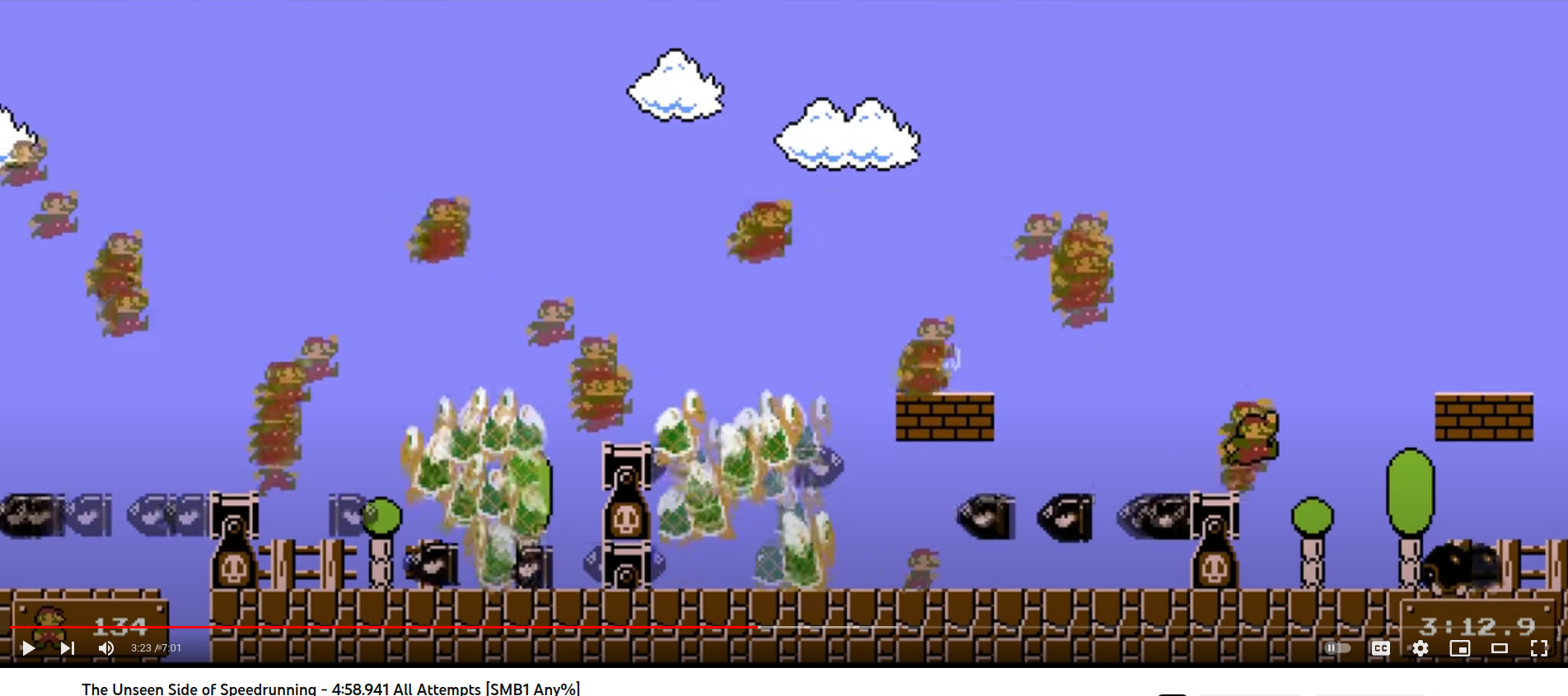}
    \caption{Screenshot from the video, ``The Unseen Side of Speedrunning - 4:58.941 All Attempts,'' 
 uploaded by FlibidyDibidy  \url{https://www.youtube.com/watch?v=X_eXSzyZudM)}}
    \label{fig:enter-label}
\end{figure}
Introducing the concept of state space as the set of points in $n$-dimensional space plus a time dimension containing the points of trajectories, we can then think of these trajectories as lying in a tube of trajectories in this space.
A trajectory in time is, therefore,
$$
(t, \gamma (t)) \in [0, + \infty) \otimes \mathbf{R}^n
$$
If the empirical data are interpreted probabilistically, it can be said that the games played lead to trajectories that have a high probability of belonging to this tube if the player is a professional. The distribution of trajectories, then, induces a potential probabilistic pattern of trajectories.  
If the game were completely deterministic and the player always played optimally, the trajectory obtained would be the one that minimized the time bounded by the category constraints.
This statement is valid as an approximation because it relies on the implicit assumption that there is a single optimal trajectory.\\
For example, for SMB1, which is a relatively simple and pseudo-deterministic game, there are different potential optimal trajectories explored empirically (using the methods of tool-assisted speedrunning \footnote{A tool-assisted speedrun is generally defined as a speedrun performed using tools that allow you to replay a set of pre-recorded, precise inputs.
There are tools designed to work with both real hardware (such as the Nintendo NES console) and emulators.}), but there is no theoretical proof or disproof of the uniqueness of the optimal trajectory, nor is there a theoretical tool for constructing an optimal trajectory.\footnote{\cite{tricky} and \cite{superai} are interesting in this regard.}\\
If we assume that the optimal trajectory exists and is known to the player, we can assume that the trajectories relative to the actual runs thicken around this trajectory.
Alternatively, we can assume that there are several equivalent optimal trajectories.   
With this premise instead, we can have a first approximation that the different runs, with different weights, remain around the different optimal trajectories.
From a classical physics perspective, the first hypothesis could be reinterpreted through variational calculus.\footnote{It would also be interesting (but outside the scope of this article) to explore the variational interpretation of d'Alembert's classical principle using the conceptual tool of virtual displacement and the implementation of collisions. For example, in many games, displacements within an object such as a wall (which are virtual in classical physics) are actually calculated by the simulator. In this case, the displacements are virtual in the simulations because a routine compensates for them, but they are real calculations in the simulator.}\\
The optimal trajectory is defined as the minimum that emerges from the principle of least action.
In this case, to make the concept consistent with the formalism, it is useful to set the time relative to every trajectory.
Thus, $t_1$ in this reasoning is the start time of the generic game within the simulation. The time $t_1$ can be taken to be formally equal to 0.
The final time $t_2$ can be taken for convenience as the maximum time of the set of games played by the player:
$$t_2=\max \{ t_i : t_i\in \textit{game time} \} $$
In this context we can write, in analogy with classical mechanics, the action functional
$$
S_{cl} = \int_{t_1}^{t_2} L \; dt
$$
where the function $L$ is in mechanics the Lagrangian function.  In conservative systems, ${\textstyle L=T-V}$, where $T$ represents the kinetic energy and $V$ the potential.
The action is related to the  trajectory of a mechanical system by the equation
$$
 \delta   S = 0
$$
More rigorously, the path taken by the system between times $t_1$ and $t_2$, and generalized configurations defined by $q_1$ and $q_2$, is the one for which the action is stationary (does not change) to first order. In the first instance, we can think of the optimal trajectory as being, in the context of speedrunning, a perfect run with respect to the constraints.\\
In the case under consideration, it is the function that incorporates the constraints and must represent the activity of the player force acting in the virtual environment. In this context, the player force could be interpreted, at least as a first approximation, as an implementation of a hypothetical potential. In this case, therefore, the forces of the simulated nature are partially implemented by the simulation (e.g., by an algorithm that prevents interpenetration of the simulated objects) and by the player themselves.
To be formally correct, we must point out that all forces, both those internal to the simulation and those induced by the player, are always mediated by algorithmic routines within the simulation.\\
We can, therefore, assume that these two classes of forces can be represented to first approximation by potentials  in the function $L$.
In this context, we thus have a first physical interpretation of a speedrunner's run. The player is the implementer of forces and the metaphysical agent (with respect to the simulation) that represents  the extremum principle.\\
The optimal player is a correct representation within the framework of classical physics. From a classical point of view, if a player deviates from the optimum, they are expressing a different Lagrangian (or constraint). In this sense, since a generic speedrunner is not optimal, the next step of our work will help understand how to keep satisfying the constraint while allowing non-optimal trajectories.
We should note that a subset of simple games can be played optimally in any run, so as we will see, the threshold between this classical framework and the next one we will present is a combination of the complexity of the game and the skill of a generic speedrunner.
 
\section{Path integrals}
\begin{figure}[H]
    \centering
    \includegraphics[width=0.5\linewidth]{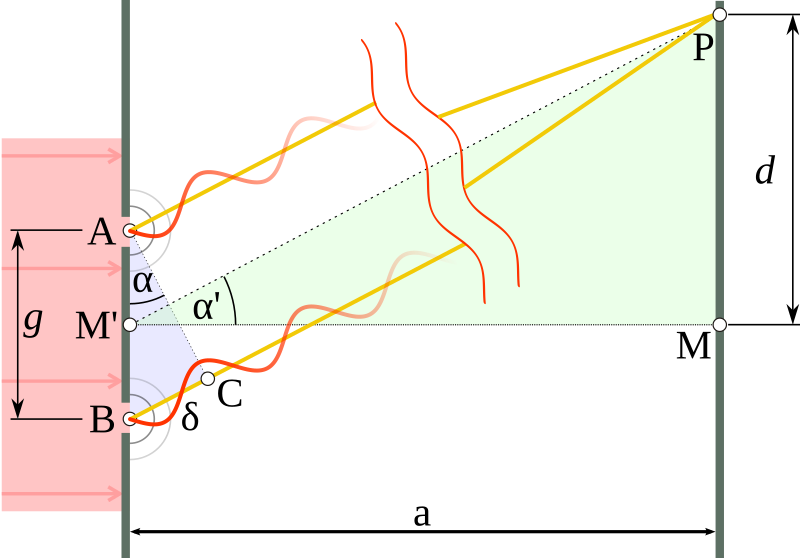}
    \caption{Double slit experiment (schematic).} 
    \label{fig:double}
\end{figure}

In the previous section, speedrunning was linked to the variational interpretation of classical mechanics.
In the next step, we would like to move on to quantum mechanics by means of the path integral.
The path integral is a technique developed by Feynman from the work of Dirac. It allows us to represent the evolution of a wave function, a central object of quantum mechanics, as the sum of infinitely many paths from a starting point in space-time to a terminal point.
It is possible to give a pseudo-intuitive interpretation starting from a fundamental topic in quantum mechanics, the double-slit experiment.
In this experiment, a quantum object, represented by a wave function impinging on a screen with two slits, can be represented after the slits as two waves that are generated (and thus add up linearly) from the slits (see figure~\ref{fig:double}). In this configuration, we therefore have two paths. The idea of the integral over the paths can be seen as a   limit transition   in which the slits and screens become infinite.
This representation is interesting for several reasons. In this text, we only explore some of its features, primarily the fact that the solution that emerges from this representation in the case of an $S$-functional (the action functional\footnote{In the context of our thesis, the point of view of \cite{action} is particularly interesting. }) with a large value compared with the Planck constant $\hbar $ tends to the classical solution. This is the solution that one would obtain with a classical particle by applying the variational calculus that was explained in the previous section.
The path integral, therefore, has classical mechanics as its limit. The limit occurs as $\hbar \rightarrow  0$.
The central idea is that each path has a weight associated with it, which makes it possible to calculate (at least theoretically) its probability.\footnote{See, for example, \cite{rattazzi}.}If we denote by $K(x_f, t_f; x_i, t_f)$ the quantum probability amplitude that an initial state $x_i, t_i$ leads to a final state $ x_i, t_f$, then in this formulation, it will take the value
$$
K(x_f, t_f; x_i, t_f) = \int_{\textit{all trajectories}} \Psi(\gamma)
$$ 
where $\gamma$ means any trajectory that has as its constraint that
$x_{\gamma}(t_i) = x_i$ e $x_{\gamma}(t_f) = x_f$.
Without going into the details of the motivation, let us point out that the weighting of these trajectories proposed by Feynman is as follows:
$$
    K(x_f, t_f; x_i, t_f) = \int_{\gamma} e^{i \frac{S[\gamma]}{\hbar}}
$$
The coefficient therefore depends on $S$, which is the function of the action.
The value $K$ does not represent the probability $P$ from transition, but is related to the probability by the extension of Born's rule \cite{born}:
$$
P = |K|^2
$$
As we have said, this representation has several peculiarities. One useful feature is that it lends itself to being studied using perturbation techniques.
That is, the action $S$ can be expanded in series by transforming the integral into a product of proxy elements of the probability product.
If we consider the series expansion of $S$, we have that the product can be decomposed into the leading term relative to the parameter determined by the classical action, which carries the initial and final state constraints, multiplied by the remaining elements, which are the quantum fluctuations relevant only when $S$ is small.\\
So, the action that represents the forces and is used to find the classical trajectory determines the individual coefficients.
It can be shown that when $S$ dominates $\hbar$, that is, in the macroscopic case, the probability of the classical trajectory tends to 1 because the other trajectories tend to cancel out. In the microscopic case, however, this result no longer holds, so to calculate $K$ it is theoretically necessary not to neglect any trajectory. In this case, the great complexity of the representation becomes apparent, with the need to use tools such as Feynman diagrams and renormalization to extract a value from the formula.
We will not go into the details of the quantum component in this text, but we will mention that the classical limit can also make sense in the type of simulation we have analyzed.
 
\section{Path integrals and speedrunning}
The next step is, therefore, to try to move from the Lagrangian representation to the path integral formulation, also in the case of the simulation or game. This step is done, at least in this work, by analogy.
If we wanted to transform the analogy into a formally consistent representation, it would be necessary to give a formal shape to the dynamics of the simulation. It is possible, at least in the first instance, to carry out this formal step by imposing further constraints and making explicit some routines for handling the virtual dynamics, but the aim of this work is to provide a more general picture, albeit by analogy.\footnote{This topic would also be an opportunity for discussion of how to overcome the difference between classical and quantum probability for the specific question \cite{classicbsquantum}.}\\
As we mentioned before, the result achieved by a perfect speedrunner is the theoretical optimal run.
For an optimal speedrunner, on the other hand, we can think of the optimal run as one that settles on a given ideal path that is not the theoretical best possible. This is due to a combination of physical and biological limitations. If we consider the optimal run, we can think that if a speedrunner's runs tend toward infinity, the neighborhood of that optimal run tends to occur almost all of the time.
Sub-optimal runs/trajectories will have a given probability of occurrence. Empirically, this probability is represented by the ratio of their frequency to the total number of runs.
This argument can be made in this simple form because, in the case of the games under consideration (we always keep the SMB1 prototype in mind), the set of possible runs is discrete, and if it is limited in time, the set becomes finite. This follows from the fact that the computer running the simulation is quantized in both time and space.\\
In time, the finiteness is represented by the concept of frames. The simulation proceeds by discrete instants which can be represented by the refresh rate of the screen. For SMB1, for example, the speed is 60 frames per second (fps).
In reality, this is itself an approximation but there is still an ultimate time limit, which is represented by the time it takes the processor to execute a single instruction.
The space is quantized because the virtual space is a subset of the memory of the simulator. The memory of a real simulator is of a finite size.
The calculation of the relative frequencies of the trajectories is, therefore, a matter of combinatorial calculus\cite{marvin}.
In physics, the transition from the classical solution to the path integral formulation introduces the quantum fact that the classical optimal trajectory is not the only one possible, given the same constraints. According to the theory, there is an infinite number of possible trajectories, weighted, as we have seen, by a given coefficient representing the possibility of realization.
This assertion is heuristically acceptable and indirectly supported by the predictive effectiveness of the theory\footnote{See, for example, \cite{precision}.} even if it cannot be demonstrated in a timely manner since one cannot empirically test the infinity of possibilities in finite time and space.\footnote{See, for example, an interpretation of Wittgenstein's antiplatonism \cite{wittengstein}. }
This point is what connects speedrunning to the representation of paths.
Speedrunning makes different paths with given probabilities actual. So again, given the same constraints (in a deterministic game), different trajectories are realized.
In general, then, one can think of representing the probability amplitude of a game as
$$
K(x_f, t_f; x_i, t_f) = \sum_{\gamma} f(S(\gamma))
$$
where, instead of making the coefficient explicit, we refer to a function $f$ of the action.
This representation is a precursor to the quantum one; the Feynman path integral seen above is a special case of it.
We can therefore relate, as in quantum mechanics, $K$ and the probability $P$ with
$$
P = |K|^2
$$
The question of what a wave function means within the context of a simulated reality, such as the one presented, is a complex topic.
However, it is possible to state that, as in quantum mechanics, a simulated game has an experiential reality that is perceived through the screen, which is based on an underlying reality that is represented by the memory registers of the simulator.\\
The reality represented on the screen is, therefore, equivalent to a measurement of the underlying reality, analogous, for example, to the pair consisting of the wave function of a quantum object and the concept of a classical physical object that can be attributed to the output of a measurement.
It is our intention to elaborate on this matter from a technical point of view in subsequent works.
In this context, however, it is interesting to note that the simulation-player point of view induces dual local and non-local contexts akin to those of quantum mechanics.
One interpretation of the quantum wave function is that it carries a non-local component of the object. For example, in the double-slit experiment, the interference pattern that generates the probability of a measurement on the screen is non-local but the measured object has local characteristics.
Similarly, from the point of view of physics, quantum mechanics has a non-local component that cannot be used to transmit information at superluminal speeds.
Regarding the simulated reality of the game, the player, who is the metaphysical agent, is capable of knowing the entire simulated universe.  \\
This translates into the ability to make decisions that are not local to the simulated world.
Metaphysical time is orthogonal to simulated time. A simulated quantum of time can have any length of metaphysical time, but has the property of being a real and finite value in metaphysical reality.
Despite this knowledge and ability, however, the player in a game such as SMB1 cannot act on it outside of the simulated physics. In practical terms, if a player knows, for example, that there is a certain item in a certain place before they start the game or before they look at the screen, they still have to obey the laws of ``classical'' physics of the game and the limits of, for example, speed in order to reach the given item. 
 
\section{Possible interpretation of the classical limit}
 
In a Feynman path integral, the parameter $\hbar$  defines the transition between classical mechanics as a limit and quantum mechanics.\footnote{Since $\hbar$ represents the  reduced Planck constant,  this consideration is almost straightforward.}
If we want to have an analogue in speedrunning, we need to correctly interpret this duality and the meaning of an equivalent threshold parameter. To interpret this threshold parameter in speedrunning, it is useful to consider how little an optimal player can deviate from the optimal path. 
This is always true on a macroscopic level. For example, even SMB1 has a microscopic level that the player attempts to control but is not entirely under their control.
The player's input can modify the simulated reality about 60 times per second in SMB1. This time interval is below the biological limit of both stimulus interpretation and muscular response.
In games like SMB1 in particular, but also in most speedrunning games in general, what is known as ``muscle memory'' plays a key role.\footnote{See, for example, \cite{posthuman}.} Physical patterns are trained to reduce conscious control time.\\
However, even in this state, there is a physical limit to accuracy, which adds up to a technical limit to the accuracy of the input tools. Input tools convert movement into an electrical signal and, therefore, have a mechanical component that reduces accuracy.
From a spatial point of view, the player also has a perception that is limited by the instrument that produces the image: the screen or a visor.
In the case of SMB1, the smallest visible element is the pixel, and the game has 256 $\times$ 240 pixels. This unit is not fundamental because the positions of objects are defined by a fraction of a pixel, the subpixel. In SMB1, there are theoretically 256 subpixels for each pixel, but these are managed in groups of 16 in the game code, so there are 16 different values for each pixel.
This fraction is manageable for professional players, but remains invisible. As in the temporal case, the more one approaches the microscopic, the more the potential randomness increases.\\
If we also assume that the player is optimal, but subject to the physiological constraints of a human body, then the optimal path cannot be followed with certainty at the microscopic level relative to the simulation. 
Thus, if the classical limit is to be interpreted as the level below which the probability with respect to the trajectory breaks down into different paths, the parameter defining the classical limit is player- and simulation-dependent.\\
The more a player is able to reduce randomness in even the smallest movements, the lower the ``quantum'' uncertainty and the threshold.
Considering that the $\hbar$ in this case depends on the skill of the player, that it makes no difference for the time of the game if two games in the real world start at very different times (they always overlap in the simulation time, as will be better explained in the next section) and that it is reasonable to assume that a game with a long history of both speedrunning and normal play will see a change over time in the type of player and the quality of the game, we can think about how $\hbar$ evolves over time.  
Over time, more and more players achieve higher performance and the number of casual players decreases. 
 In this scenario, therefore, the randomness decreases and so does the threshold of the ``classical limit.''
 The $\hbar$ becomes smaller and the ``quantum world'' of the game recedes.

\section{Quantum interpretations and the many worlds theory}
The mathematical formalism of quantum mechanics lends itself to various philosophical interpretations. The intrinsic probability in the quantum context, mediated by Born's rule, is one of the most important factors driving the philosophical discourse. That is, the interpretation is often extended from the quantum case to physical reality in general. For example, the Copenhagen interpretation, first proposed by the physicist Niels Bohr in 1920 and de facto the mainstream interpretation, ascribes ontological value to the probabilistic component; the quantum object that translates into physical reality is probabilistic in nature.
There are other elements of quantum mechanics that undergo this process of interpretation, such as wave--particle duality, the reality of particles in relation to the concept of fields, the concepts of time and causality in relation to the reversibility of the equations, and the concept of locality.
There is also the distinction between real and virtual particles. A virtual particle can be interpreted both as a mathematical device in quantum field theory and as a transient particle.\\
With regard to the particle concept, it is useful to ask, in the context of a simulation or game such as the one analyzed, what kind of element can be interpreted as a particle. 
The avatar, for example, can be seen as a particle that has a given speed and position only when it is observed (i.e., 60 times per second in SMB1 when it is on the screen).
In the simulator, the position and speed of an object in the simulation are the result of certain memory allocations and some algorithmic components. \\
It is interesting to note briefly that other objects that have these characteristics (i.e., that are represented in memory by comparable values in equivalent allocations to those of a real object) and that interact with the same algorithms or parts of the same algorithms, becoming invisible agents on the dynamics of real objects, are indeed pseudo-particles or virtual particles in this context.\\
Two major alternatives to the Copenhagen interpretation are the de Broglie--Bohm theory and the many worlds theory (MWT).
The latter is probably the one that, if properly reinterpreted, allows an interesting evaluation of the proposed analogy.
The aforementioned advantages of studying the simulated-player game model are that the simulated physics is completely under the control of a human operator, and that there is a component of the metaphysics of the simulated system (i.e., the physical reality of the player) that is amenable to investigation.\\
Instead of postulating, as the Copenhagen interpretation does, that probability in quantum mechanics is ontological and cannot be further investigated, MWT interprets it as the separation of universes that are in superposition.
In the famous example of Schrödinger's cat \cite{cat}, according to MWT, the two states of the cat are in superposition until the act of measurement. After the measurement, the universe splits into two almost identical copies, one in which the cat is alive and one in which it is dead.
At least as far as classical quantum theory is concerned, the two interpretations are empirically indistinguishable because they do not innovate the mathematical component and do not predict different phenomena.\\
Even if a speedrunner's games are not superimposed in the player's physical space, they are clearly superimposed in simulated space.
This is because they start at the same simulated time and in the same space.
In terms of these dimensions, two games in which the path taken is identical for a given interval of time, but differs at the end of that interval, can be interpreted as a pair of universes that separate after a measurement.

\section{Path integral and Machine learning in speedrunning}

A formalisation of player activity as a force field can also be used to provide a common layer between different optimisation and machine learning techniques.
As mentioned above, the use of tools to search for optimal strategies is now central to the speedrunning community  and various machine learning techniques have been applied to this topic.
Returning to the \cite{superai} and \cite{tricky} references, it can be seen that the problem of finding an algorithmic optimisation tool to support speedrunning is a complex one.\\
Part of the complexity comes from the fact that, in the absence of a common theoretical layer, the approach to optimisation is usually highly dependent on the characteristics of the individual game.\\
Despite the diversity of game types, there is one unifying element: the speedrunner. The speedrunner is a unifying element in seeking the best result. Human beings are indeed capable of playing very different games in their quest for the best time or performance.
This starting point allows for the idea that a mathematical formalisation of the speedrunner can be used to generalise machine learning and AI methods.\\
It should also be noted that in the case of techniques that use multiple algorithmic agents to optimise an objective function \footnote{there is a large literature dealing with multi-agent reinforcement learning techniques, see for example \cite{multiagent}}, the human playing a game can be interpreted as an algorithmic agent itself.

$$
\textit{Agents} \rightarrow \textit{path integral representation} \rightarrow \textit{Game-simulation}
$$

In this context, an abstract layer between the player and the game can help to integrate the actions of different agents from a theoretical point of view. \\
By being able to represent, even partially, the player's action by an action function, it is also possible to use available machine learning tools to find numerical solutions to problems already addressed in quantum physics \footnote{See for example \cite{learningintegrals}}.\\
An interesting addition to an abstract theory introduced by speedrunning is that optimal strategies must be feasible for a human to implement \footnote{An interesting video analysing the human limit in playing SMB was posted on YouTube by Summoning Salt \url{https://www.youtube.com/watch?v=7rIJNT7dCmE}}.
This new constraint brings the search for the optimal path in speedrunning closer to the search for the optimum in a neural network. 
In the case of neural networks, the constraint is  made explicit through training. The training set  represents a set of problems solved by the human expert that must be replicated and optimised.\\
There are also different studies that attempt to use a representation inspired by quantum field theory to solve problems and gain insights in the context of neural networks \footnote{See for example \cite{neural}}.
In this area too, it is necessary to try to find a common ground that can identify similarities between techniques that at first sight appear to be very different.

\section{Conclusions}
In this article, we have tried to draw a parallel between quantum mechanics and speedrunning.
In both contexts, the analogy is a potential vehicle for philosophical speculation. With respect to the former, it can be used to reason about the problem of a possible metaphysical context around the physical world. The advantage here is to have a toy model in which the metaphysical component and the physical world are explicable, and in which interactions can be analyzed.\\
In particular, it is possible to provide an interpretation of two fundamental concepts: the stationary phase and ontological probability. 
In the context of speedrunning, on the other hand, a possible formalized representation of the action performed by the player in the act of playing can be introduced.
The analogy can also be further formalized. For this, a necessary step is the creation of a toy model of the algorithmic component that handles the physical simulation of a game.
We have also shown a possible way to make effective use of such a formalisation in the context of using machine learning algorithms.\\
We also believe that this analogy, albeit for a more limited purpose, can be, in certain contexts,  more malleable for speculative analysis than conjectures such as the many worlds theory or even the simulation hypothesis in its modern incarnation by Nick Bostrom \cite{nick}.

\end{document}